\documentclass[iop,apj]{emulateapj}

\usepackage{amsmath}
\usepackage{mathrsfs}

\journalinfo{ApJ, 633:474-488, 2005 November 1 --- arXiv:1002.3325v1 [astro-ph.SR]}
\submitted{Received 2004 August 5; accepted 2005 May 16}
\shorttitle{Slow Solar Wind}
\shortauthors{Rappazzo et al.}

\begin{document}


\title{Diamagnetic and Expansion Effects on the Observable Properties \\ of the Slow Solar Wind in a Coronal Streamer}



\author{A. F. Rappazzo\altaffilmark{1}}
\affil{Dipartimento di Fisica, Universit\`a di Pisa, 56127 Pisa, Italy;
rappazzo@df.unipi.it}

\author{M. Velli\altaffilmark{2}}
\affil{Jet Propulsion Laboratory, Pasadena, CA 91109; 
marco.velli@jpl.nasa.gov}

\author{G. Einaudi\altaffilmark{3}}
\affil{Dipartimento di Fisica, Universit\`a di Pisa, 56127 Pisa, Italy;
einaudi@df.unipi.it}


\author{R. B. Dahlburg}
\affil{Laboratory for Computational Physics \& Fluid Dynamics, \\ Naval Research Laboratory, Washington, DC 20375-5344, USA; 
rdahlbur@lcp.nrl.navy.mil}


\altaffiltext{1}{Also at Berkeley Research Associates, Springfield, VA 22150}

\altaffiltext{2}{Dipartimento di Astronomia e Scienza dello Spazio,  \\ Universit\`a di Firenze, 50125 Florence, Italy}

\altaffiltext{3}{Also at Department of Physics and Astronomy, George Mason University, Fairfax, VA 22030}



\begin{abstract}
The plasma density enhancements recently observed by the Large-Angle 
Spectrometric Coronagraph (LASCO) instrument onboard the Solar and
Heliospheric Observatory (SOHO) spacecraft have sparked considerable 
interest. In our previous theoretical study of the formation and initial motion of 
these density enhancements
it is found that beyond the helmet cusp of a coronal
streamer the magnetized wake configuration is resistively unstable, that a 
traveling magnetic island
develops at the center of the streamer, and that density enhancements occur within the magnetic islands. As the massive magnetic island travels outward, both its speed and width increase. The island passively traces the acceleration of the inner part of the wake.

In the present paper a few spherical geometry effects are included, taking into account either the radial divergence of the magnetic field lines and the average expansion suffered by a parcel of plasma propagating outward, using the Expanding Box Model (EBM), and the diamagnetic force due to the overall magnetic field radial gradients, the so-called
melon-seed force.
It is found that the values of the acceleration and density contrasts
can be in good agreement with LASCO observations, provided the spherical divergence of the magnetic lines starts beyond a critical distance from the Sun and the initial stage of the formation and acceleration of the plasmoid is due to the cartesian evolution of MHD instabilities. This result provides a constraint on the topology of the magnetic field in the coronal streamer.
\end{abstract}





\section{Introduction}

\begin{figure*}[t]
  \includegraphics[width=.9\textwidth]{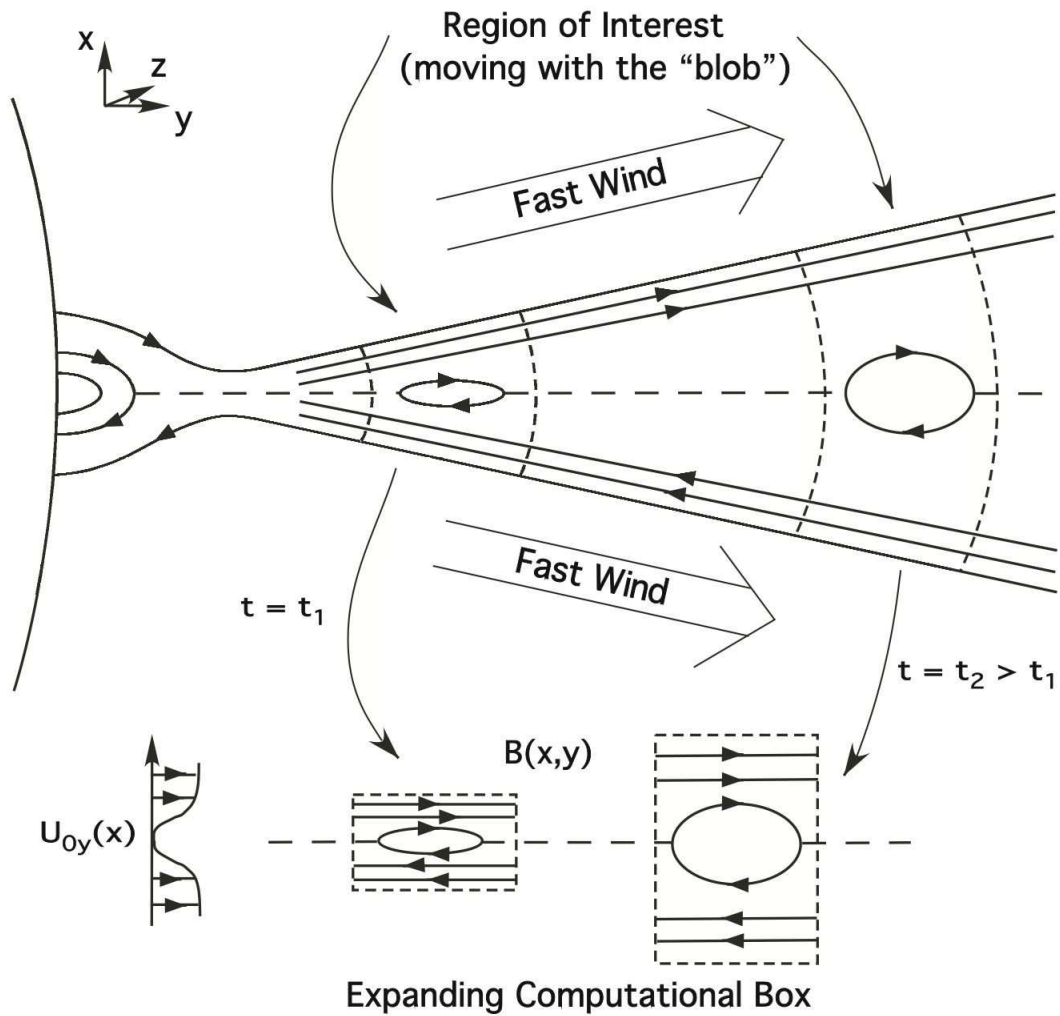}  
  \caption{\footnotesize Expanding Box Model: coordinate system and notations. The spatial coordinate aligned with the 
  mean flow (streamwise direction) is denoted by $y$, $x$ is the spatial coordinate along which the mean flow varies 
  (cross-stream direction) and $z$ is the coordinate along which physical fields are invariant (spanwise direction).   We 
  approximated the region of interest by a parallelepiped  whose lengths are $L$ along $y$, $a$ and $b$ along $x$ and 
  $z$ respectively (not drawn in figure). The distance from the center of the Sun is $R(t)$, which changes with time when 
  the plasmoid moves outward. }
  \label{fig:fig1}
\end{figure*}
One of the most interesting findings of the LASCO instrument onboard
the \emph{Solar and Heliospheric Observatory} (SOHO) spacecraft has
been the observation of a continuous outflow of material in the solar
streamer belt. An analysis performed using a difference image
technique (\citet{sheeley:1997}, \citet{wang:1998}) has revealed the 
presence of plasma
density enhancements, called ``blobs'', accelerating away
from the Sun.  These plasmoids are seen to originate just beyond the
cusps of helmet streamers as radially elongated structures a few
percent denser than the surrounding plasma sheet, of approximately $1
R_{\odot}$ in length and $0.1 R_{\odot}$ in width.  They are observed
to accelerate radially outward maintaining constant angular spans at a
nearly constant acceleration up to the velocity of $200$--$450 \: km
\, s^{-1}$, in the spatial region between about $5$ and $30
R_{\odot}$.  It has been inferred that the blobs are ``tracers'' of
the slow wind, being carried out by the ambient plasma flow 
\citep{sheeley:1997}.

The solar streamer belt is a structure consisting of a magnetic 
configuration centered on the current sheet, which extends above the 
cusp of a coronal helmet.  The region underlying the cusp is made up 
of closed magnetic structures, with the cusp representing the point 
where separatrices between closed and open field lines intersect.  
Further from the Sun, at solar minimum, the streamer belt around the 
equator appears as a laminar configuration consisting of a thick 
plasma sheet with a density about $1$ order of magnitude higher than 
the surrounding plasma, in which much narrower and complex structures 
are embedded.  As a first approximation, moving from the center of the 
streamer in polar directions at radii greater than the radius of the 
cusp, the radial component of the magnetic field increases from zero, 
having opposite values on the two sides of the current sheet.  
As far as the flow distribution 
is concerned, the fast solar wind originates from the unipolar regions 
outside the streamer belt, while the slowest flows are located at the 
center of the sheet.

Explaining how both the slow and the fast component of the solar wind 
are accelerated is one of the outstanding problem in solar physics.  
Although the association between the slow solar wind and the streamer 
belt (e.g. \citet{gosling}) and between the fast wind and the polar coronal holes is broadly 
recognized, the mechanism which leads to such accelerations is still 
a matter of debate. \citet{einaudi:1999} developed a magnetohydrodynamic 
model (later extended to the compressible case in \citet{ein:2001}) that
accounts for many of the typical features observed in the slow
component of the solar wind: the region above the cusp of a helmet
streamer is modelled as a current sheet embedded in a broader wake
flow.   This and previous studies (\citet{dbe:1998}, \citet{dahl:1998})
 show that reconnection of the magnetic field
occurs at the current sheet and that in the non-linear regime, when
the equilibrium magnetic field is substantially modified, a
Kelvin-Helmoltz instability develops, leading to the
acceleration of density enhanced magnetic islands \citep{ein:2001}.

The problem we study has  a roughly spherical symmetry, but in our
simulations we approximate the region beyond the cusp of a helmet
streamer with a rectangular box, and the geometry of the fields with a
cartesian geometry, instead of a spherical one.  To perform the
numerical calculations we use a 2D numerical code with periodic
boundary conditions in the streamwise direction (the spatial direction 
aligned with the mean flow) and non-reflecting
boundary conditions in the cross-stream direction (the spatial direction
along which the initial mean flow varies).  A key feature, at
the same time an advantage and a limitation of our analysis, is that
thanks to the periodic boundary conditions that we use in the
direction of the flow we are allowed to proceed in all our numerical
simulations in a pseudo-Lagrangian fashion: i.e.  the size of the
computional box is fixed and then the time evolution of the spatial
region moving with the box is followed in time, simply because what
goes out from one side comes in from the other.  In this case we
follow the development of the magnetic island during its acceleration
outward along the radial direction. With respect to spatially developing studies of the solar wind our time developing method is able to reach a higher
resolution. In our case in fact the computational domain  is smaller and we  are allowed to increase the grid resolution along the current sheet in order to
 resolve the  small scales developing during the non-linear stage, and hence
enabling us to study the development of a resistive-like instability
that would be lost in a spatially developing study with a much lower resolution.
On the other hand, so far no one has been able to study this phenomenon in
spherical geometry, and we are planning to do it as the next step in the 
development of our model.

In the present paper we introduce two major effects related to the 
spherical symmetry of the problem we study. In what follows we suppose that
when the magnetic island, the ``blob'', moves outward along the 
radial direction it suffers a spherical expansion due to the approximate 
spherical topology of the physical fields (Figure~\ref{fig:fig1}).
Although the subsequent  algebra is quite complicated, as this approach enables us to attach the problem thanks to the availability of enough grid points, we still want to use the previous pseudo-Lagrangian time developing  technique.  This necessarily requires the use of a rectangular computational box and fields with a Cartesian geometry, but we can include the expansion suffered by a parcel of plasma propagating outward by making the box expand at the 
correct rate. This is done by inserting a suitable force term in the momentum 
equation, a generalization to the variable velocity regime of the Expanding 
Box Model developed by \citet{grap:1996}, \citet{grap:1993}, as described 
in the following sections. The other spherical effect included in the present 
paper is the so-called ``melon-seed'' force. This is a diamagnetic force that 
a region of closed magnetic field lines suffers when it is embedded in a 
radially divergent background magnetic field (see \citet{park:1954}, 
\citet{park:1957} and \citet{sc:2000}). While in a spherical geometry this force
would rise naturally, in our Cartesian approximation we take it into account by 
inserting a suitable force term in the momentum equation, in analogy with the 
Expanding Box Model. 

Section~\ref{sec:pm} describes the governing equations as well as the initial and boundary conditions, the last ones described with more details in the 
Appendix. In section~\ref{sec:re} we detail our numerical results. 
Section~\ref{sec:disc} contains a discussion of our results, as well as some 
concluding remarks.

\section{Physical Model}    \label{sec:pm}

\subsection{Governing Equations} \label{sec:ge}

In our analysis and simulations we use the compressible, isothermal, 
dissipative, three dimensional magnetohydrodynamic (MHD) equations, 
that we write here in dimensionless form:
\begin{equation} \label{eq:mhd1}
\frac{\partial \rho}{\partial t} +\boldsymbol{ \nabla } \cdot \left( \rho \boldsymbol{u} \right) = 0,
\end{equation}
\begin{eqnarray}
\rho \left( \frac{\partial {\boldsymbol{u}}}{\partial t} + \boldsymbol{u} \cdot \boldsymbol{\nabla} \boldsymbol{u} \right) = - 
\boldsymbol{\nabla} \left( P + \frac{\boldsymbol{B}^2}{2} \right) & &+ \boldsymbol{B} \cdot 
\boldsymbol{\nabla} \boldsymbol{B} \qquad \quad \\
& & \quad  + \frac{1}{\mathcal R} \boldsymbol{\nabla} \cdot \boldsymbol{\xi}, \label{eq:mhd2}
\end{eqnarray}
\begin{equation} \label{eq:mhd3}
\frac{\partial {\boldsymbol{B}}}{\partial t} = \boldsymbol{\nabla} \times \left( \boldsymbol{u} \times \boldsymbol{B} \right) + 
\frac{1}{\mathcal{R}_m} \nabla^2 \boldsymbol{B},
\end{equation}
supplemented by the equation of state $P = \rho T$.  In the above 
equations $\rho (\boldsymbol {x},t)$ is the mass density, $\boldsymbol{u} (\boldsymbol{x}, 
t)$ is the flow velocity, $P (\boldsymbol{x}, t)$ is the thermal pressure, 
$\boldsymbol{B} (\boldsymbol{x}, t)$ is the magnetic induction field, $T (\boldsymbol{x}, 
t)$ is the plasma temperature and $\xi_{ij}= \frac{\partial u_i}{\partial x_j} + 
\frac{\partial u_j}{\partial x_i} - \frac{2}{3} \delta_{ij} \boldsymbol{\nabla} \cdot 
\boldsymbol{u}$ is the viscous stress tensor.  To render non-dimensional the 
equations we used the characteristic quantities $L^{\ast}$, 
$u^{\ast}$, $\rho^{\ast}$, and the related quantities $t^{\ast}$, 
$T^{\ast}$, $B^{\ast}$
\begin{equation} 
t^{\ast} = \frac{L^{\ast}}{u^{\ast}}, \quad \ T^{\ast} = m_{p} \left( 
u^{\ast} \right)^2, \quad \ B^{\ast} = u^{\ast} \sqrt{ 4\pi 
\rho^{\ast} },
\end{equation}
where $m_{p}$ is the proton mass.  We use a simplified diffusion model, 
in which both the magnetic 
resistivity ($\eta$) and the shear viscosity ($\mu$) are constant and 
uniform. A Stokes relationship is assumed, so that for the bulk 
viscosity  $\lambda = \left( \frac{2}{3} \right) \mu $. The 
kinetic and magnetic Reynolds numbers~$\mathcal{R}$ and 
$\mathcal{R}_m$ are then given by:
\begin{equation}
\mathcal{R} = \frac{\rho^{\ast} u^{\ast} L^{\ast}}{\mu},
\qquad \mathcal{R}_m = \frac{4 \pi u^{\ast} L^{\ast}}{\eta c^2}.
\end{equation}

To ensure the solenoidality of the magnetic field, and supposing that there is no variation of the fields along $z$, we introduce the magnetic potential $\phi$ defined by:
\begin{equation}
\boldsymbol{B} = \boldsymbol{\nabla} \times \left( \phi \, \boldsymbol{e}_z \right) + B_z \, \boldsymbol{e}_z,
\end{equation}
and replace equation~(\ref{eq:mhd3}) with:{\setlength\arraycolsep{-1em}
\begin{eqnarray} 
&& \frac{\partial \phi}{\partial t} = -u_x \frac{\partial \phi}{\partial x} -u_y \frac{\partial \phi}{\partial y} +
\frac{1}{\mathcal{R}_m} \left[ \frac{\partial^2 \phi}{\partial x^2} + \frac{\partial^2 \phi}{\partial y^2} \right]
\\ \label{eq:mhd4} 
& &\frac{\partial B_z}{\partial t} =   - B_z \left( \frac{\partial u_x}{\partial x} 
+ \frac{\partial u_y}{\partial y} \right)  + B_x \frac{\partial u_z}{\partial x} 
+ B_y \frac{\partial u_z}{\partial y} \\
&& \qquad \quad
 - u_x \frac{\partial B_z}{\partial x}
- u_y \frac{\partial B_z}{\partial y} + \frac{1}{\mathcal{R}_m} \left[ 
\frac{\partial^2 B_z}{\partial x^2} + \frac{\partial^2 B_z}{\partial y^2} \right], \label{eq:mhd5}
\end{eqnarray}
}

The equations solved in our numerical simulations depend on the problem that is studied: equations \mbox{(\ref{eq:mhd1})-(\ref{eq:mhd2})}, \mbox{(\ref{eq:mhd4})-(\ref{eq:mhd5})} are solved when we do not take into account the expansion and the diamagnetic force, whereas when we consider these last effects some new force terms are added to the equations \mbox{(\ref{eq:mhd1})-(\ref{eq:mhd2})}, \mbox{(\ref{eq:mhd4})-(\ref{eq:mhd5})}  and  a coordinate transformation is implemented for a matter of convenience, as described forward in this section. In any case the numerical problem is simplified by assuming that there is no variation in one of the spatial coordinates ($z$).  In the following we refer to the spatial coordinate aligned with the mean flow as the streamwise direction~($y$), the spatial coordinate along which the mean flow varies as the cross-stream direction~($x$), and to the remaining (spanwise) direction as~$z$.  The system has periodic boundary conditions in the streamwise direction, along which a Fourier pseudospectral method is used in the numerical computations, and nonreflecting boundary conditions, achieved via the method of projected characteristics (\citet{thom:1987}, \citet{thom:1990}, \citet{vana:1989}), in the cross-stream direction, where a compact finite difference scheme of the sixth order is used \citep{lele:1992}. 
This method, coupled with a hyperbolic tangent mesh stretching around the current sheet, allows reasonable Reynolds numbers to be achieved without increasing the number of grid points dramatically.  We solve the equations in a 2D box whose dimensions are $l_{x} \times l_{y} = \pm 10.63 \times 2 \pi / \alpha$, where $\alpha$ is the streamwise wavenumber (the value of the box length along the $x$ direction is the result of  the mesh stretching).  Time is discretized with a third-order Runge-Kutta method.  In the simulations that we present in this paper we used a numerical grid  with $n_{x} \times n_{y} = 401 \times 256$ points, and for all simulations the Reynolds numbers are $\mathcal{R} = \mathcal{R}_m = 200$.  For a better description of the numerical techniques used in our code see \citet{ein:2001}.

The region where the acceleration of the solar wind occurs would be more aptly described in spherical coordinates, but in order to perform our numerical simulations we do a Cartesian approximation, i.e. we approximate the fields topology with a Cartesian one. In this way we can use our pseudo-Lagrangian technique using a 2D code. As previously said we include the spherical expansion suffered by a parcel of plasma propagating outward by making expand our computational box at the correct rate. The region we study (see Figure~\ref{fig:fig1}) is a  spherical sector of angular extent $\alpha$ in the cross-stream direction ($x$) and $\beta$ in the span-wise direction, whose length is $L$ along the stream-wise  direction ($y$) and whose distance from the center of the Sun is $R(t)$, where $R$ changes in time while the plasma flows outward along the radial direction. We approximate this region with a parallelepiped whose cross lengths are respectively $a$ and $b$ along the directions $x$ and $z$, and whose length in the radial direction is $L$.  We suppose that when the plasma present in this region moves outward it suffers an average expansion that we approximate with the expansion of the box (see Figure~\ref{fig:fig1}). When following the evolution in the comoving frame of reference a non-inertial force term arises.  In the Appendix it is shown that the force field for unity of mass that inserted in the momentum equation yields the correct expansion of the computational box is the following:
\begin{equation} \label{eq:fexp0}
\boldsymbol{\mathit{f}}_{exp} (\boldsymbol{x},t) = \frac{\ddot{R}(t)}{R(t)} \Big( x, 0, z \Big).
\end{equation}
In the previous equation the velocity of the box and its
distance from the center of the sun are defined as:
\begin{equation} \label{eq:boxv}
\dot{R} (t)= \frac{1}{L_y} \int_0^{L_y} u_y(0,y,0,t) \mathrm{d} y,
\end{equation}
\begin{equation} \label{eq:boxr}
R(t) = R_0 + \int_0^t \dot{R} (t') \mathrm{d} t',
\end{equation}
i.e. we define the velocity of the box as the average velocity in the $y$ direction along the center of the computational box.  The basic numerical algorithm was developed and tested by \citet{grap:1993}, \citet{grap:1996} who showed that  the EBM gives a correct description of the evolution of waves (exact wave-action conservation) and turbulence in an expanding medium as well as capturing basic features of the expanding solar wind such as the formation of corotating interaction regions and forward and reverse shocks between high speed and low speed solar wind. The EBM does not conserve energy because in an expanding flow work is done by the plasma. Angular momentum and magnetic flux are conserved exactly however (see \citet{grap:1996} for a detailed description and tests of the physical model).  The force term~(\ref{eq:fexp0})  introduces a variation along the $x$ and $z$~coordinates, but this explicit coordinate dependence  cancels when we make a coordinate transformation  to a self-similarly expanding transverse coordinate  $\boldsymbol{x}' \left( \boldsymbol{x},t \right)$ defined by
\begin{equation}
x' = \frac{R_0}{R(t)} x, \qquad y' = y, \qquad z' =  \frac{R_0}{R(t)} z.
\end{equation}
It can be easily verified that in this way the expanding computational domain coordinate is transformed to a fixed computational grid. The equations solved by our 2D numerical code and all the mathematical details are extensively explained in the Appendix.

When we make our Cartesian approximation we approximate the radially diverging magnetic field lines with straight ones, losing the radial nonhomogeneity of the solar magnetic field. An effect that is lost with this approximation is  the so-called ``melon seed'' force, a diamagnetic force acting on a region of closed magnetic field lines. It is known (see \citet{sc:2000}, \citet{pc:1985}, \citet{park:1954} and \citet{park:1957})  that a region of closed magnetic field lines embedded in a background radially diverging magnetic field is subject to such force. Modeling this region as a prolate spheroid of volume~$V$ elongated in the radial direction it can be demonstrated that the value of the total force acting on the spheroid is approximately:
\begin{equation} \label{ms}
\boldsymbol{F}_m = - V \left[ \frac{ \mathrm{d} } {\mathrm{d} R} \left( \frac{B_e^2}{8\pi} \right) \right] \boldsymbol{\hat{e}}_y,
\end{equation}
where  $B_e = B_{e0} \left( \frac{R_0}{R} \right)^2$ is the background magnetic field, $B_{e0}$ being constant. In our model axial invariance is assumed, i.e. there is no variation in the $z$ direction in the Cartesian approximation. Hence the prolate spheroid is modeled as  a cylinder with an elliptical cross section, and the same force for unit  volume is used.  The force given in equation~(\ref{ms}) is the total force acting on this plasmoid. The force per unit  volume derived from this equation and used in our computations is given by:
\begin{equation} \label{msfl}
\boldsymbol{\mathit{f}}_m (x,y,t) = \Theta(x,y,t) \frac{B_{e0}^2}{2\pi R_0} \left( \frac{R_0}{R} \right)^5 \boldsymbol{\hat{e}}_y,
\end{equation}
that  written in non dimensional form is given by:
\begin{equation} \label{msfl2}
\boldsymbol{\mathit{f}}_m (x,y,t) = \Theta(x,y,t) \frac{2 B_{e0}^2}{R_0} \left( \frac{R_0}{R} \right)^5 \boldsymbol{\hat{e}}_y,
\end{equation}
in which $\Theta$ is a field whose value is $\Theta =1$ inside magnetic island and $\Theta =0$ outside. In this way the field~(\ref{msfl}) integrated over the volume of the plasmoid gives as a result the total force~(\ref{ms}).

The region above the cusp of a helmet streamer is assumed to be in local hydrodynamic equilibrium, therefore gravity, magnetic curvature forces and the flow all balance out. Once density fluctuations arise, the gravitational force reappears as an Archimedes buoyancy force acting on  regions with a different density compared to the average. This force field can be  written in non-dimensional form as
\begin{equation} \label{grav}
f_G =  - k_G \frac{\Delta \rho}{R^2} \quad \text{where} \quad k_G = \frac{G M_\odot}{L^\ast \left( u^\ast \right)^2 } \, , 
\end{equation}
$G$ being the Gravitational constant and $M_\odot$ the solar mass. Taking as characteristic length and velocity, the width of the wake $L^\ast = a_V \sim 0.1 R_\odot$ and the velocity of the fast solar wind $u^\ast = u_\infty \sim 6\cdot 10^5\, m\, s^{-1}$, we obtain the value $k_G =5.3$. We can now compare the relative strength of this buoyancy force with that of the melon-seed force (\ref{msfl2}). As shown in the next paragraph $A \sim 2.5$, $R_0 = 60$ (i.e. $6\,  R_\odot$ in conventional units, and supposing $\Delta \rho = 1$, we obtain near the cusp of the helmet streamer ($R = R_0$):
\begin{equation}
\frac{f_m}{f_G} \sim 
\frac{  \frac{2 A^2}{R_0}}{k_G \frac{\Delta \rho}{R^2}} \sim 100\, ,
\end{equation}
i.e. the buoyancy force is negligible with respect to the melon seed force near the cusp of the helmet streamer and in the region of interest of the simulations presented in the following paragraphs. 

In the next section we introduce as initial conditions the fields (\ref{eq:equi1})-(\ref{eq:equi2}). These fields are a one-dimensional equilibrium for the region beyond the cusp of a helmet streamer, i.e. the variation of the fields is only along the cross-stream direction (the $x$ coordinate) and Solar Corona Reynolds numbers are so high that they are substantially an equilibrium solution of equations (\ref{eq:mhd1})-(\ref{eq:mhd4}) over the time scales of our simulations. But in our computational domain Reynolds numbers are limited by resolution to values around $\mathcal{R} = \mathcal{R}_m = 200$, values which are so small that unphysical diffusion of equilibrium fields would take place. In the foregoing 
equations the pattern of the diffusion terms is
\begin{equation} \label{eq:diff}
\frac{\partial f}{\partial t} = \ldots +\frac{1}{\mathcal R} \frac{\partial^2 f}{\partial x^2},
\end{equation}
where $f$ is a generic field. As the variation of the equilibrium fields is only along the $x$~coordinate, i.e. as  a Fourier series in the $y$~direction they have only a $k=0$~mode, we avoid such an unphysical diffusion using instead of~eq.~(\ref{eq:diff}) the following one:
\begin{equation}
\frac{\partial f}{\partial t} = \ldots +\frac{1}{\mathcal R} \frac{\partial^2}{\partial x^2} \left( f
- \hat{f}_0 \right).
\end{equation}
In this way the modes~$k=0$ of the Fourier series in the $y$ direction, that we indicate with~$\hat{f}_0$, do not diffuse. This could have potentially given rise to development of excessively small scales in $x$ on the $k=0$ harmonic but empirically  the numerical filtering used along this direction has been enough.

\subsection{Initial Conditions}

We model a planar section of the solar streamer belt, in the region beyond the cusp of a helmet streamer, as a magnetic current sheet of thickness $a_B$ embedded in the center of a broader wake flow of thickness $a_V$. The plasma flows at the speed of the fast solar wind at the edges and at a much lower velocity at the sheet.  Taking as characteristic length the thickness of the wake $L^{\ast} = a_V$, and as characteristic density and velocity those of the fast wind $\rho^{\ast} = \rho_{\infty}$, $u^{\ast}=u_{\infty}$, the basic fields, written in non-dimensional form, are then given by:
\begin{equation} \label{eq:equi1}
u_{0y} (x) =1 - {\rm sech} \left( x \right),
\end{equation}
\begin{equation}
B_{0y} (x) = A \tanh \left( \delta x \right),
\qquad
B_{0z} (x) = A \ {\rm sech} \left( \delta x \right),
\end{equation}
\begin{equation} \label{eq:equi2}
\rho = 1, \qquad T = \frac{1}{M^2},
\end{equation}
where $\delta = \frac{a_V}{a_B}$ is the ratio of the two widths, and $A$ and $M$ are respectively the Alfv\'en number, the ratio between the Alfv\'en speed (${c_A}_{\infty}$) and the flow speed ($u_{\infty}$), and the sonic Mach number, the ratio between the flow speed ($u_{\infty}$) and the sound speed ($c_{s\infty}$), of the system at the edges, i.e.  of the fast wind:
\begin{equation}
A = \frac{B_{\infty}}{\sqrt{4\pi \rho_{\infty}} u_{\infty}} =
\frac{{c_A}_{\infty}}{u_{\infty}}, 
\qquad 
M = \frac{u_\infty}{\sqrt{T_\infty}}  =
\frac{u_\infty}{c_{s\infty}}.
\end{equation}
Our wake-current sheet model is then characterized by three parameters, $A$, $M$, and $\delta$, which vary considerably in the solar corona as a function of the distance from the Sun.  In this paper we are interested in studying the time
\begin{figure}
  \includegraphics[width=0.47\textwidth]{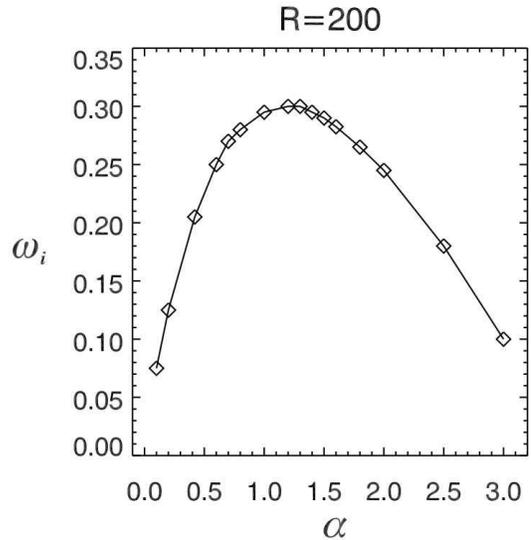}  
  \caption{\footnotesize Dispersion relation for the resistive mode with $A=1.5$, $R=200$. The growth rate ($\omega_i$) as a function of wavenumber ($\alpha$) is displayed (in non-dimensional units).}
  \label{fig:fig2}
\end{figure}
and spatial evolution of a parcel of plasma starting from the region immediately beyond the cusp of a helmet streamer.  In this region the typical Alfv\'en velocity  
\begin{figure*}
  \includegraphics[width=1.\textwidth]{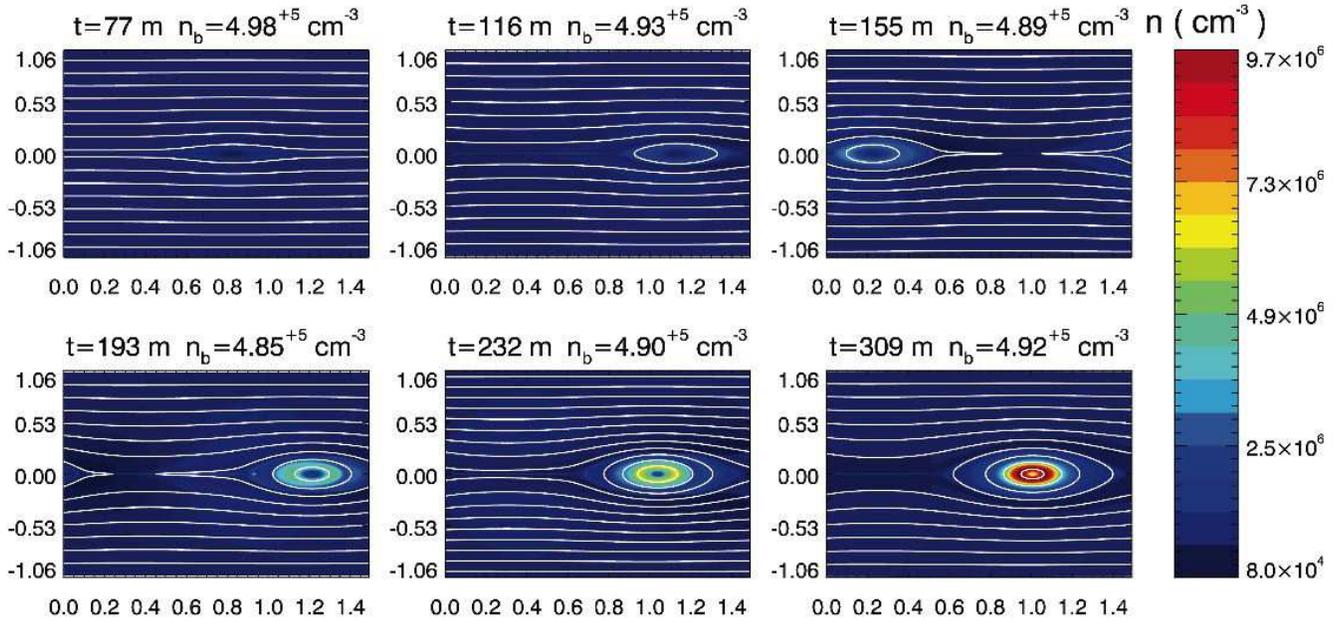}  
  \caption{\footnotesize Run~A --- No expansion or melon-seed force is included in this simulation. Time evolution of magnetic islands is showed. Magnetic field lines (\emph{white lines}) and electron density filled contours (colours show the value of the density field in accordance with the colour scale) at selected times show density enhanced magnetic islands forming and accelerating outward. The $x$ and $z$ axis are respectively along the vertical and horizontal direction and are measured in solar radii units ($R_{\odot}$). $n_b$ is the average electron density at the top and bottom $z$ boundaries ($x=\pm 1.06\, R_{\odot}$), it doesn't change significantly because the box is not expanding. A $\sim 5$ hours time should be added to the times showed to keep count of the rescaled linear time-scale. } 
 \label{fig:fig3}
\end{figure*}
\begin{figure}
  \includegraphics[width=0.47\textwidth]{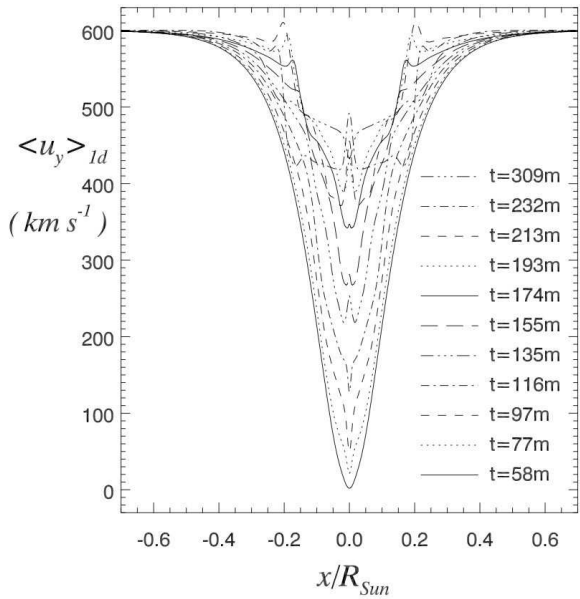}  
  \caption{\footnotesize Run~A --- Streamwise component of the velocity $(u_y)$ averaged in
   the $y$ direction as a function of $x$ (the cross-stream direction) at selected times. When the nonlinear regime 
   develops (after $t \sim 58\, m$), while magnetic reconnection continues to occur, a Kelvin-Helmholtz-like instability is 
   triggered, leading to the acceleration of the central part of the wake up to the value $\sim 500 \: km \, s^{-1}$.  }
  \label{fig:fig4}
\end{figure}
${c_A}_{\infty}$ is approximately $1000$--$1500 \: km \, s^{-1}$, the sound speed $c_{s\infty}$ is about $100 \: km \, s^{-1}$, and the velocity of the fast solar wind $u_\infty$ is $\sim 600 \: km \, s^{-1}$.  It follows that $A = {c_A}_{\infty} / {u_{\infty}} \sim 1500 / 600 = 2.5$, and $ M = u_\infty / c_{s\infty} \sim 600 / 100 = 6$.  The plasma $\beta$ is also a useful quantity, defined as $\beta = 2p / | \boldsymbol{B} |^{2} = 
2{c_{s\infty}}^{2} / {{c_A}_{\infty}}^{2} = 2 / \left( A M \right)^{2}$.  In this case we have $\beta \sim 0.0089$.  In our numerical calculations we use values of these parameters that are close to these estimates but computationally more accessible.  Thus, in our runs $A = 1.5$ and $M=2$ (hence $\beta = 2 / 9 = 0.22$). Unfortunately there is no direct observation of the heliospheric current sheet close to the Sun ($ r < 10 R_{\odot}$), hence there is no measurement of the parameter $\delta$. But from the topology of both the magnetic and the velocity fields it is reasonable to suppose $\delta >> 1$, i.e. the width of the wake flow much bigger than that of the current sheet (see \citet{stribling:1996}, \citet{einaudi:1999}). Such a high  value of $\delta$  is not computationally accessible and on the other hand we have seen in \citet{einaudi:1999} that the impact of the value of $\delta$ is mainly to change the critical value of $A$ below which the dynamics is dominated by the flow. Therefore in this paper we have chosen $\delta = 5$, which corresponds to a magnetically dominated regime. 

\begin{figure*}
  \includegraphics[width=1.\textwidth]{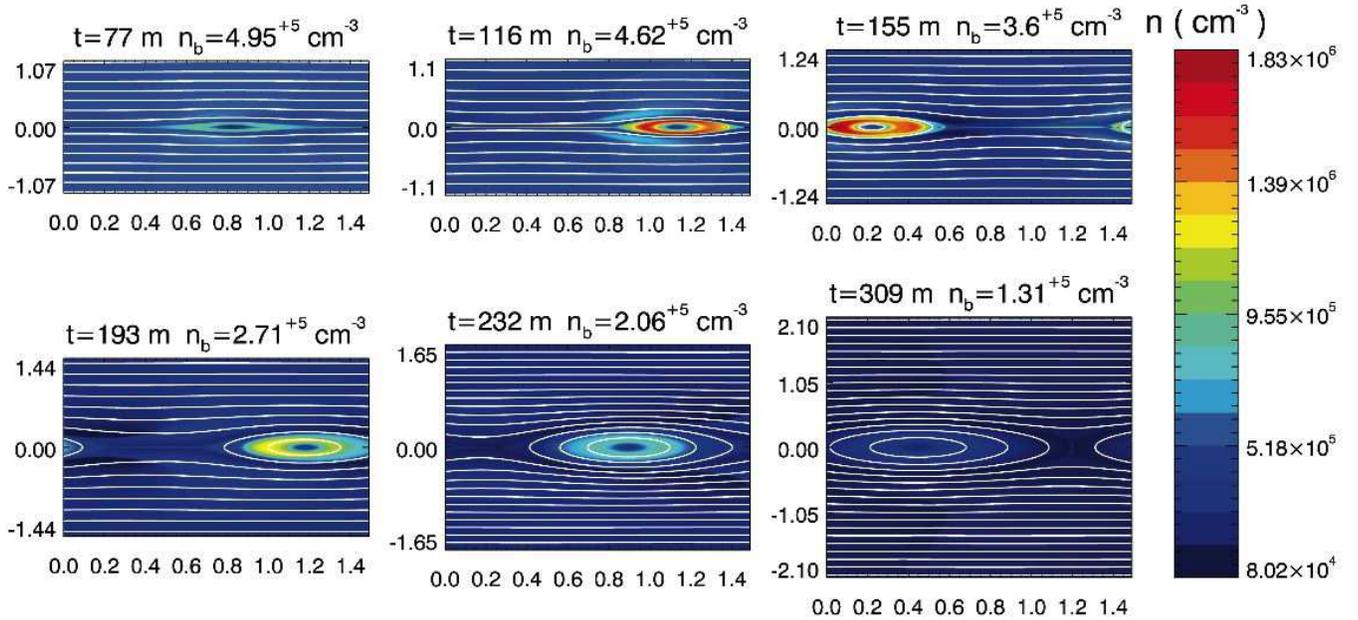}  
  \caption{\footnotesize  Run~B --- Expansion but no melon-seed force included in this simulation. Time evolution of 
  magnetic islands is showed. Magnetic field lines (\emph{white lines}) and electron density filled contours (colours show 
  the value of the density in accordance with the colour scale) at selected times show density enhanced magnetic islands 
  forming and accelerating outward.  The $x$ and $z$ axis are respectively along the vertical and horizontal direction and 
  are measured in solar radii units ($R_{\odot}$). The initial distance between the box and the center of the Sun ($R_0$) is 
  $6\, R_{\odot}$, during time evolution this distance increases, and at the selected times showed in the Figure the values 
  of $R(t)$ are given by: $6.02, 6.2, 7.01, 8.11, 9.32, 11.86\, R_{\odot}$.   $n_b$ is the average electron density at the top 
  and bottom $z$ boundaries, as the box expands its value decreases as $(R_0/R)^2$.  A $\sim 5$ hours time should be 
  added to the times showed to keep count of the rescaled linear time-scale. }
  \label{fig:fig5}
\end{figure*}

\section{Results}   \label{sec:re}

This section presents the results of our numerical simulations. In \S\ref{sec:runA} we show for reference the result of a numerical simulation in which both the melon seed force and the spherical expansion are neglected. In \S\ref{sec:runB}, in order to show the influence of the expansion on the dynamics of the wake-current sheet system, we present the results of a numerical simulation in which the geometrical expansion is implemented while the melon seed force is neglected. At last, in \S\ref{sec:runC} we show the results of a simulation in which both the melon seed force and the spherical  expansion are implemented.

If there was no magnetic field ($A=0$) the wake-current sheet system (equations~(\ref{eq:equi1})--(\ref{eq:equi2})) would be unstable for a Kelvin-Helmholtz-like instability. Two unstable modes exist in this limit: a varicose (sausage-like) mode and a sinuous (kink-like) mode. The stabilizing effect of a magnetic field parallel to the flow over the Kelvin-Helmholtz instability is well known (e.g. \citet{chandra}), and in fact our linear studies (\citet{dbe:1998}, \citet{einaudi:1999}) have found that increasing the value of the Alfv\'en number $A$ beyond a critical value the two ideal modes are stabilized. On the other hand increasing the strength of the magnetic field leads the system to be unstable to a tearing-like instability. This resistive varicose  mode has the same spatial symmetry of the varicose fluid mode and,  while in the linear regime the dynamics is not substantially modified by the presence of the flow, developing like a classic tearing instability, in the non-linear regime the presence of the flow has  a dramatic influence. In fact, as the perturbed magnetic and velocity fields attain finite amplitude, the resulting nonlinear stresses along the magnetic island borders \citep{dahl:1998} lead to a transfer of momentum between the flow and the magnetic island. As a result the magnetic island and the central part of the wake are accelerated.
The coupling between magnetic reconnection and Kelvin-Helmholtz instability that develops in the dynamics of this mode has been studied in \citep{einaudi:1999}.

In all calculations a small amplitude perturbation of the resistive varicose type (see   \citet{dbe:1998}) at the wavenumber $\alpha = 0.42$ was added to the initial state. Although our numerical code was developed to study the non-linear dynamics, we can estimate (via a fitting during the linear regime) the growth rates of the perturbations at various wavenumbers. Figure~\ref{fig:fig2} shows the dispersion relation for resistive instabilities obtained with this method. It is evident from this Figure that we have not chosen the fastest growing  wavenumber for our nonlinear calculations. The reason is related to the coupling of the magnetic reconnection instability and of the Kelvin-Helmholtz instability.  The initiation of the Kelvin-Helmholtz instability occurs when the 
resistive instability is well into its nonlinear regime.  The maximum desired value for $\alpha$ is limited by the 
well known 
fact  (e.g. \citet{ray:1982}, \citet{ferr:1983}) that the
Kelvin-Helmholtz instability is stabilized for wavelengths approaching the thickness of the shear layer (in our case the amplitude of the wake flow~$a_V$), in this way the wavelengths that give rise to a more efficient acceleration of the wake are the longer ones (compared to the amplitude of the wake).  The minimum desired value for $\alpha$ is determined by the number of subharmonics we want to include, to allow for vortex merging.  For this reason, based on the Kelvin-Helmholtz dynamics, we have picked up the value $\alpha =0.42$ and not the wavenumber of the fastest growing resistive mode.  A most comprehensive study of this mechanism is underway, in particular a multiwavelength one.

While the numerical computations have been carried out using 
non-dimensional equations, for a matter of clarity the results will be 
presented in the ordinary dimensional units. As in this paper we are interested in studying the time and[ spatial evolution of a parcel of plasma starting from the region immediately beyond the cusp of a helmet streamer, i.e. $4-6 \ R_{\odot}$,
 we have chosen as characteristic quantities those characterizing this region.
 We have chosen as characteristic length the thickness of the wake $L^{\ast} = 0.1 R_{\odot}$, as characteristic velocity that of the fast wind $u^{\ast} = 600 \: km \, s^{-1}$, and as characteristic density the current sheet electron density  at $\sim 5 R_{\odot}$, i.e. $n = 5 \cdot 10^5 \: cm^{-3}$ \citep{guha:1996}.
 From these quantities we derive the characteristic time $t^{\ast} = L^{\ast}/u^{\ast}  \sim 116 \, s$. A particular caution should be taken when considering the times presented in the following simulations. Our numerical simulations have been in fact performed using Reynolds numbers values ($R=200$)  much lower than coronal values. While most of the nonlinear dynamics is due to ideal processes 
\begin{figure}
  \includegraphics[width=0.47\textwidth]{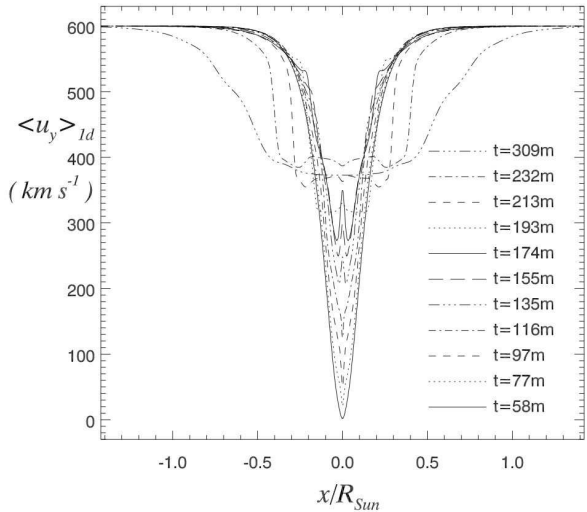}  
  \caption{\footnotesize Run~B --- Streamwise component of the velocity $(u_y)$ averaged in
   the $y$ direction as a function of $x$ (the cross-stream direction) at selected times. Expansion effects imply a 
   weakening of cross-stream gradients of magnetic and velocity fields, leading to a lower acceleration profile.}
  \label{fig:fig6}
\end{figure}
(i.e. Kelvin-Helmholtz instability)
 which are little affected by dissipation, the growth rate of the tearing-like instability developing in the linear regime is strongly affected by the Lundquist number, hence the time-scale of the linear stage must be rescaled. With a Lundquist number of the order $S=10^9$ the linear phase should last approximately 5 hours \citep{einaudi:1999}. This time should be added to the times of the simulations presented in the following
 paragraphs.

\subsection{Run A---No expansion or melon-seed force}   \label{sec:runA}

In this simulation we consider neither the expansion nor the diamagnetic force. While the parameters are somewhat different, we reproduce the essential evolutionary features reported by \citet{einaudi:1999} and 
\citet{ein:2001}. Figure~\ref{fig:fig3} shows the formation and acceleration of density enhanced magnetic islands.  These magnetic islands form with streamwise length equal to the
perturbation wavelength.  The magnetic island cross-stream width grows from
a small amplitude in the linear regime to the order of $a_B = 0.02 \, R_{\odot}$, the width of the current sheet, at the beginning of the non-linear stage ($t \sim 77m$).  While magnetic reconnection continues to occur in the nonlinear regime, the Kelvin-Helmholtz instability also is triggered, leading to the acceleration of the central part of the wake (Figure~\ref{fig:fig4}) up to the value $\sim 500 \: km \, s^{-1}$ at $t=309m$

\begin{figure}
  \includegraphics[width=0.4\textwidth]{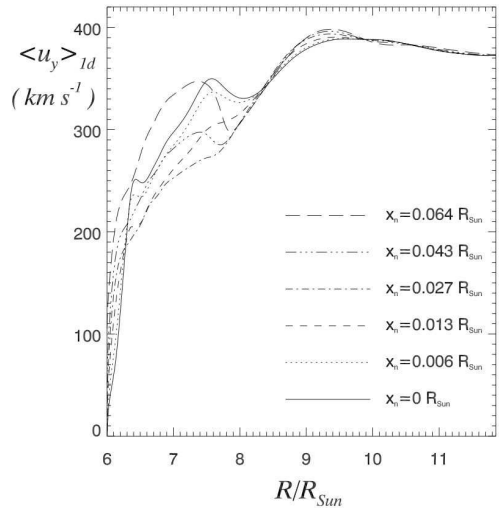}  
  \caption{\footnotesize Run~B --- Average streamwise component of the velocity ($<u_{y}>$) 
  at selected $x_n$ values ($x_n$ representing the non-expanding numerical 
  grid related to the expanding physical grid $x$ by the relation 
  $x = x_n \cdot R(t)/R_0$) as a function of $R$, the distance of the expanding 
  box from the center of the Sun.}
  \label{fig:fig7}
\end{figure}

\begin{figure*}
  \includegraphics[width=1.\textwidth]{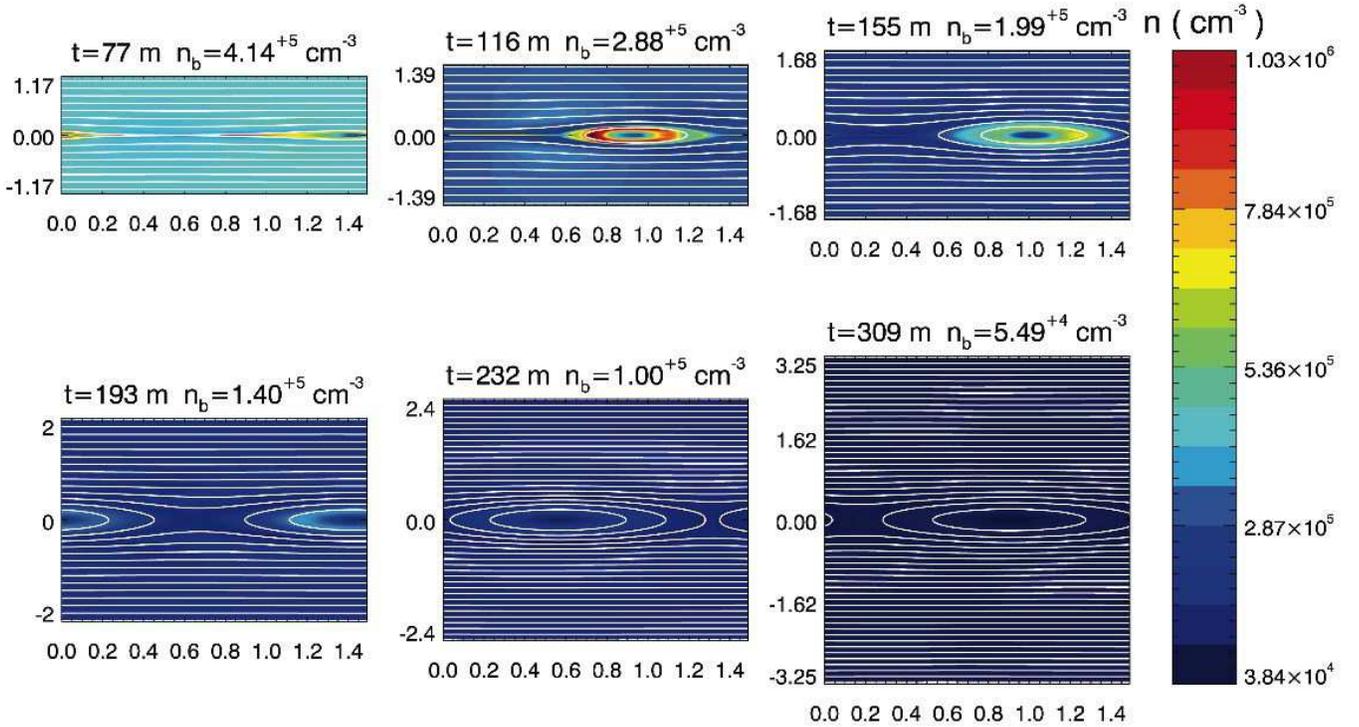}  
  \caption{\footnotesize Run~C --- Expansion and melon-seed force implemented. Time evolution of magnetic islands is showed. Magnetic field lines (\emph{white lines}) and electron density filled contours (colours show the value of the  density in accordance with the colour scale) at selected times are showed.
 The value of the melon-seed force is very small during the linear stage,
 but as non-linear stage occurs, both Kelvin-Helmholtz and the diamagnetic
 force accelerate the island. 
The $x$ and $z$ axis are respectively along the vertical and horizontal direction 
and are measured in solar radii units ($R_{\odot}$). The initial distance between 
the box and the center of the Sun ($R_0$) is $6\, R_{\odot}$, during time 
evolution this distance increases, and at the selected times showed in the 
Figure the values of $R(t)$ are given by: $6.58, 7.82, 9.46, 11.4, 13.5, 18.3\, 
R_{\odot}$.   $n_b$ is the average electron density at the top and bottom $z$ 
boundaries, as the box expands its value decreases as $(R_0/R)^2$. A $\sim 
5$ hours time should be added to the times showed to keep count of the 
rescaled linear time-scale. } 
  \label{fig:fig8}
\end{figure*}

The value of the density enhancement ($n_{max}/n_b$ defined as the ratio of 
the maximum density inside the magnetic island over the background density 
value, steadily increases from an initial uniform
 value ($n_{max} = n_b =5 \cdot 10^5 \: cm^{-3}$) up to a value which is roughly $20$ times greater ($n_{max} \sim  9.7\cdot 10^6$) at $t=309m$. Whereas  the formation and evolution of this density enhanced magnetic island is very similar to that of the observed ``blobs" (i.e. we have a dense, accelerated plasmoid), both the density enhancement and the acceleration profile result in larger values than the ones observed (\citet{sheeley:1997},  \citet{wang:1998}).   This has been one of the main reasons to study the more realistic configurations described in the next two sections. We will show in the next section that when expansion effects are included such a peaked density does not occur because, as expected,  while the magnetic island travels outward there is a rarefaction of the island due to the expansion.

\subsection{Run B---Expansion but no melon-seed force}   \label{sec:runB}

This run provides us with information on the effects of the geometrical expansion on the dynamics of the wake-current sheet model for slow solar wind formation and acceleration.  In this section we neglect the diamagnetic force.  We will account for this force in the next section.

We used as starting radius $R_0 = 6 R_{\odot}$, approximately
the radius of the region located near the cusp of a helmet streamer. During the time evolution the radius $R$ increases its value up to $R\sim 12
 \: R_{\odot}$ at $t=309m$. From Figure~\ref{fig:fig1} we can clearly notice that while the plasmoid moves outward the box suffers an expansion. Its volume
 increases as $\left( R/R_0 \right)^2$ leading to an average decreasing
 of the mass density as $\left( R_0/R \right)^2$, as the values of the background
 density $n_b$ indicated in Figure~\ref{fig:fig1} show.

The dynamical evolution (Figure~\ref{fig:fig5}) for the expansion case exhibits some crucial differences from that described  in the previous section for the non-expanding case.  
Again we have the formation of a density enhanced magnetic island, but as now 
expansion is occurring the average density of matter decreases as $\left( R_0/R
\right)^2$, and the density enhancement is lower. The density enhancement $n_{max}/n_b$ for the selected times shown in Figure~\ref{fig:fig5} range between $2.5$ and $5$.  This value is closer, with respect to the non-expanding
simulation presented in the previous paragraph,  to the observed one, which is 
of the order of $7 \pm 2 \%$ of the background $K$ corona \citep{sheeley:1997}.  
\citet{dahl:1998} showed that the transfer of energy to the perturbed fields from 
the background magnetic and velocity fields depends strongly on the 
cross-stream gradients of these fields. On the other hand the expansion of 
the box implies flow velocities orthogonal to the radial direction which lead to a 
weakening of these field's gradients \citep{grap:1996} and as a consequence to 
a lower acceleration profile, shown in Figure~\ref{fig:fig6}. 

In figure~\ref{fig:fig7} we plot the average streamwise velocity at
selected $x_n$ values (here $x_n$ is the non-expanding numerical grid, related to the expanding physical grid $x$ by the relation $x = x_n \cdot R(t)/R_0$) as a function of the
distance between the box and the center of the Sun, which reaches the
value  $R \sim 12 R_{\odot}$ in conventional units,
at $t \sim 309m$. 
This velocity profile is in good quantitative agreement with the one obtained by observations (compare with Fig.~4 on page~L166 of \citet{wang:1998}).

\subsection{Run C---Expansion and melon-seed force}   \label{sec:runC}

In this section we present the results of a numerical simulation in which both the geometric expansion and the diamagnetic, or melon seed, force are implemented.

As detailed at the end of \S\ref{sec:ge} the region above the cusp of a helmet streamer is assumed to be in local hydrodynamic equilibrium, therefore gravity, magnetic curvature forces and the flow all balance out. Once density fluctuations arise, the gravitational force reappears as an Archimedes buoyancy force acting on regions with a different density compared to the average. We have shown that this buoyancy force is negligible with respect to the melon-seed force in the region of interest of our simulation.
\begin{figure}
  \includegraphics[width=0.47\textwidth]{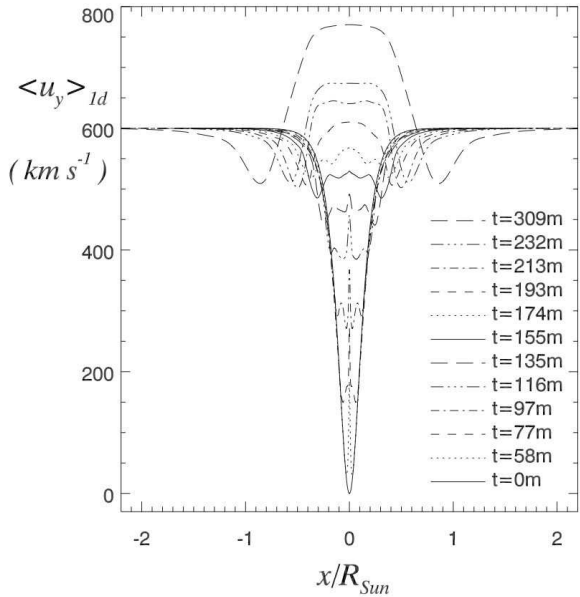}  
  \caption{\footnotesize Run~C --- Streamwise component of the velocity $(u_y)$ averaged in
   the $y$ direction as a function of $x$ (the cross-stream function)  at selected times.  The value of the melon-seed force is very small during the linear stage,
 but as non-linear stage occurs, both Kelvin-Helmholtz and the diamagnetic
 force synergistically act to accelerate the magnetic island leading to an over-acceleration}
  \label{fig:fig9}
\end{figure}

\begin{figure}
  \includegraphics[width=0.47\textwidth]{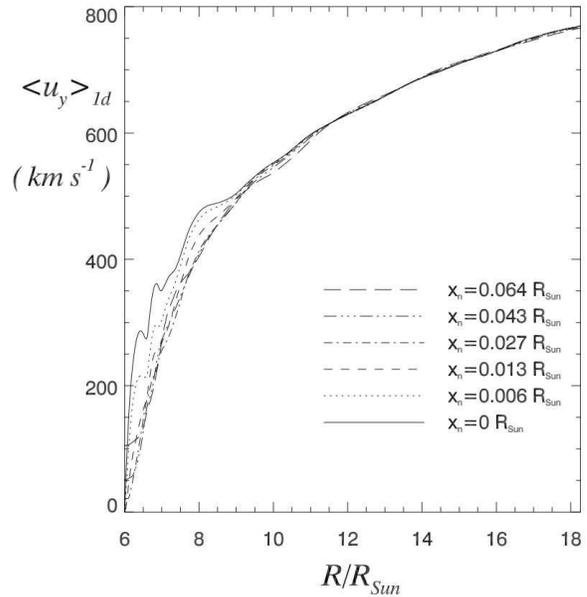}  
  \caption{\footnotesize Run~C --- Average streamwise component of the velocity ($<u_{y}>$) 
 at selected $x_n$ values ($x_n$ representing the non-expanding numerical grid related to the expanding physical grid $x$ by the relation $x = x_n \cdot R(t)/R_0$) as a function of $R$, the distance of the expanding box from the center of the Sun.}
  \label{fig:fig10}
\end{figure}

The expression for the melon seed force~(\ref{msfl2}) holds for a magnetic 
island embedded in an external magnetic field $B_e = B_{e0} \left( \frac{R_0}{R} 
\right)^2$. We compute the value of $B_e$ as the value of the magnetic field
at the outer border of the magnetic island. Hence the value of the melon-seed 
force is very small during the linear stage, when the magnetic island is smaller
than the current sheet width and the ambient magnetic field is small too.
Figure~\ref{fig:fig8} shows the time evolution of the system. During the linear regime the diamagnetic force is negligible, so that the dynamical evolution is
very similar to the ones presented in the previous simulations. But as the
non-linear stage occurs, both the Kelvin-Helmholtz and the melon-seed force,
synergistically accelerate the island, leading to an over-acceleration of the 
plasmoid, as shown in Figures~\ref{fig:fig9}--\ref{fig:fig10}. 

On the other hand, the expression of the diamagnetic force~(\ref{msfl2}) holds for a plasmoid in a radial divergent background magnetic field. Indeed we have implemented this diamagnetic force supposing that the magnetic field in the
region just beyond the cusp of an helmet streamer diverges radially. 
Unfortunately there is no direct observation of the topology of the magnetic field in this region. But as the resulting acceleration profile is not observed (\citet{sheeley:1997}, \citet{wang:1998})  we conclude from the results of this simulation that in such a region the magnetic field line do not diverge (in other words, the melon seed must be much smaller than the semi-empirical estimate used in our simulations). Both more refined observations of this region and theoretical studies, including high resolution simulations in a fully spherical geometry, are desirable to establish what kind of magnetic topology we have in this region. 

\section{Discussion}   \label{sec:disc}

In this paper, we have refined the model presented in \citet{einaudi:1999}, \citet{ein:2001}, for the formation and acceleration of the slow component of the solar wind within a coronal streamer. In particular, we have been concerned with the study of the initial stage of the process.  The main results of the present paper can be summarized as follows:

\begin{enumerate}

\item The expanding box generalization of the wake-current sheet model exhibits the formation and outward acceleration of density enhanced magnetic islands. The plasmoid width, length and acceleration profiles are in good agreement with LASCO observations.

\item The geometrical expansion by itself has a stabilizing effect on the Kelvin-Helmholtz instability developed in the non-linear regime.  It also is responsible for a rarefaction of the accelerated plasmoid.

\item From our simulations there is a strong indication that the topology of the magnetic field lines in the region close but beyond the cusp of an helmet streamer is not radially divergent.  This divergence would lead to as yet unobserved acceleration and density profiles due to the diamagnetic force. Note however that other effects (i.e. drag from the denser current sheets) might be important as well. 

\end{enumerate}

In conclusion, these computations reinforce the scenario that the structure and acceleration of the slow solar wind can be attributed primarily to the intrinsic instability and subsequent evolution of the underlying wake-current sheet system. To reproduce the observational details properly, further theoretical work and additional solar data are needed. From the theoretical point of view, we need a study of more complex and realistic basic configurations, where an initial density profile peaked on the current sheet is present, and a non force-free magnetic field is chosen.

\acknowledgments

A.F.R. thanks C.~Calonaci of Cineca~(Bologna) for his help in optimizing the numerical code. The numerical computations were performed on the SGI ORIGIN~3000 in Cineca thanks to an agreement with CINECA-CNAA. A.F.R., M.V., and G.E. were supported by the Ministero Universit\`a Ricerca Scientifica Tecnologica (MURST) contract COFIN n.~2002025872. A.F.R. and M.V. were
also supported  by the European network TOSTISP contract 
n.~HPRN-CT-2001-00310. R.B.D. was supported by the NASA Sun Earth Connection Guest Investigator Program and the NASA Sun Earth Connection Theory Program. A.F.R. and M.V. thank the IPAM program ``Grand Challenge
Problems in Computational Astrophysics'' at UCLA where this paper was
completed. We would also like to thank the referee for the stimulating and 
fruitful discussion about the physical model presented in this paper.






\appendix  \label{sec:app}

As previously said we suppose that when the magnetic island moves outward the plasma present in the region in and around it suffers a spherical expansion due to the approximate spherical topology of the physical fields (Figure~\ref{fig:fig1}).
When a particle or a fluid element moves outward along the radial coordinate $r$ at fixed azimuthal  and polar angles, the ratios $x/y$ and $z/y$ are constant in time. For the sake of simplicity consider a system of spherical coordinates where $\theta$ is the polar angle from the y-axis (with $ 0 \le \theta \le \pi $),  $\phi$ is the azimuthal angle in the zx-plane from the  z-axis (with $ 0 \le \phi < 2 \pi$), and $r$ is the distance from the origin. As usual the relations between the spherical and the cartesian coordinates are given by:
\begin{equation}
y=r \cos \theta , \qquad z=r \sin \theta \cos \phi , \qquad  x=r \sin \theta \sin \phi.
\end{equation}
It can be easily verified that the ratios $x/y$ and $z/y$ depend only by the angles $\theta$ and $\phi$; hence for a particle or a fluid element moving along the radial coordinate these ratios are constant in time. In our case $y=R(t)$ and for such a fluid element we can write:
\begin{equation} \label{eq:exp}
\frac{x(t)}{R(t)}=\frac{x_0}{R_0}, \qquad \frac{z(t)}{R(t)}=\frac{z_0}{R_0},
\end{equation}
where $x_0$, $z_0$ and $R_0$ are the initial values of these quantities.  Differentiating with respect to time we have 
\begin{equation}
\dot{x}(t) = \frac{x_0}{R_0} \dot{R}(t) =  \frac{x(t)}{R(t)} \dot{R}(t)  , \qquad \dot{z} (t) = \frac{z_0}{R_0} \dot{R}(t) =   \frac{z(t)}{R(t)} \dot{R}(t).
\end{equation}
This means that the cross-component of the velocity is given by:
\begin{equation}
\boldsymbol {u}_{\perp} = \frac{\dot{R} (t)}{R(t)} \bigg( x(t), 0, z(t) \bigg).
\end{equation}
This holds for a particle or a fluid element moving outward along the radial coordinate. To apply this to our subject we replace this velocity with the fluid velocity and look for a force able to cause this fluid motion. In other words, in our picture a fluid that is expanding spherically is characterized by a velocity field whose cross-component is given by:
\begin{equation}
\boldsymbol {u}_{\perp}(\boldsymbol{x},t) = \frac{\dot{R} (t)}{R(t)} \bigg( x, 0, z \bigg).
\end{equation}
Hence we introduce a new force in the momentum equation corresponding to the spherical expansion effects, i.e. the force term whose solution is given by the previous term. To find this force term we simply insert in the momentum equation the velocity field $\boldsymbol{u}_{\perp}(\boldsymbol{x},t)$ which gives us the force field for unity of mass
\begin{equation} \label{eq:fexp}
\boldsymbol{\mathit{f}}_{exp} (\boldsymbol{x},t) = \frac{\partial \boldsymbol{u}_{\perp}}{\partial t} + \left( \boldsymbol{u}_{\perp} \cdot \boldsymbol{\nabla}  \right) \boldsymbol{u}_{\perp} = \frac{\ddot{R}(t)}{R(t)} \Big( x, 0, z \Big).
\end{equation}
We have chosen to define the velocity of the box $\dot{R}$ as the average velocity in the $y$ direction along the center of the computational box (see eq.~\ref{eq:boxv}), and its distance from the center of the Sun $R$ accordingly (see eq.~\ref{eq:boxr}).  

In order to perform the numerical simulations the computational box must be fixed. For this reason we consider the coordinate transformation $\boldsymbol{x}' \left( \boldsymbol{x},t \right)$ defined by:
\begin{equation}
x' = \frac{R_0}{R(t)} x, \qquad y' = y, \qquad z' =  \frac{R_0}{R(t)} z.
\end{equation}
It can be easily verified that this coordinate transformations brings the physical dominium expanding with the law~(\ref{eq:exp}) into a fixed one. This coordinate transformations implies a transformations of the field. To the field $f(\boldsymbol{x},t)$ corresponds a new field $g(\boldsymbol{x'},t)$ defined by
\begin{equation}
f(\boldsymbol{x},t) = g (\boldsymbol{x'}(\boldsymbol{x},t)).
\end{equation}
The following relations can be easily verified:
\begin{equation}
\frac{\partial f}{\partial t}  (\boldsymbol{x},t) = \frac{\partial g}{\partial t} (\boldsymbol{x}'(\boldsymbol{x},t),t) - \left[ \boldsymbol{U} (\boldsymbol{x}'(\boldsymbol{x},t),t) \cdot \boldsymbol{ \nabla '} \right] g (\boldsymbol{x}' (\boldsymbol{x},t),t)
\end{equation}
\begin{equation}
\boldsymbol{ \nabla} f (\boldsymbol{x},t) = \boldsymbol{ \nabla '} g (\boldsymbol{x}'(\boldsymbol{x},t),t)
\end{equation}
where
\begin{equation}
\boldsymbol{U}(\boldsymbol{x'},t)=\left( \frac{\dot{R}(t)}{R_0} x', 0,\frac{\dot{R}(t)}{R_0} z' \right) = \frac{\dot{R}(t)}{R_0} \left( x', 0, z' \right) 
\end{equation}
\begin{equation}
\boldsymbol{ \nabla'} = \left( \frac{R_0}{R(t)} \frac{\partial}{\partial x'}, \frac{\partial}{\partial y'},
\frac{R_0}{R(t)} \frac{\partial}{\partial z'} \right)
\end{equation}
Defining the transformed fields $\rho$, $T$, $\boldsymbol u$, $\boldsymbol B$, with a primed index, like for example
\begin{equation}
\boldsymbol{u} (\boldsymbol{x},t) = \boldsymbol{u'} (\boldsymbol{x'}(\boldsymbol{x},t)),
\end{equation}
and noticing that the transformed expansion force field~(\ref{eq:fexp}) has the following form
\begin{equation} \label{eq:exprim}
\boldsymbol{\mathit{f'}}_{exp} (\boldsymbol{x'},t) = \frac{\ddot{R}(t)}{R(t)} \Big( x', 0, z' \Big) = \frac{ \partial \boldsymbol{U} }{\partial t} (\boldsymbol{x'},t),
\end{equation}
the MHD equations (for clarity we explicitly include the isothermal equation) can be written in the new coordinate system and with the transformed fields as:
\begin{equation}
\frac{\partial \rho'}{\partial t} = \left( \boldsymbol{U} \cdot \boldsymbol{ \nabla' } \right) \rho' - \boldsymbol{ \nabla' } \cdot \left( \rho' \boldsymbol{u'} \right)
\end{equation}
\begin{equation}
\frac{\partial \boldsymbol{u'}}{\partial t} = \left( \boldsymbol{U} \cdot \boldsymbol{\nabla'} \right) \boldsymbol{u'} + \boldsymbol{\mathit{f'}}_{exp} - \left( \boldsymbol{u'} \cdot \boldsymbol{\nabla'} \right) \boldsymbol{u'}
-\frac{1}{\rho'} \left[ \boldsymbol{ \nabla' } \left( P' + \frac{\boldsymbol{B'}^2}{2} \right) - \left( \boldsymbol{B'} \cdot \boldsymbol{ \nabla' } \right) \boldsymbol{B'} \right]  
+ \frac{1}{\rho' \mathcal{R}} \boldsymbol{ \nabla' } \cdot \xi'
\end{equation}
\begin{equation}
\frac{\partial \boldsymbol{B'}}{\partial t}  = \left( \boldsymbol{U} \cdot \boldsymbol{ \nabla' } \right) \boldsymbol{B'} + 
\boldsymbol{ \nabla' } \times \left( \boldsymbol{u'} \times \boldsymbol{B'} \right) + \frac{1}{\mathcal{R}_m} \nabla'^2 \boldsymbol{B'}
\end{equation}
\begin{equation}
\frac{\partial T'}{\partial t} = \left( \boldsymbol{U} \cdot \boldsymbol{ \nabla' } \right) T' - \left( \boldsymbol{u'} \cdot \boldsymbol{ \nabla' } \right) T'
\end{equation}
where
\begin{equation}
\xi'_{ij}= e'_{ij} -\frac{2}{3} \boldsymbol{ \nabla' } \cdot \boldsymbol{u'} \delta_{ij},
\qquad e'_{ij}=\partial'_i u'_j + \partial'_j u'_i, \qquad P'= \rho' T'.
\end{equation}
In order to simplify the previous equation notice that
\begin{equation}
\boldsymbol{\nabla'} \cdot \boldsymbol{U} = 2~\frac{\dot R}{R}~, \qquad \left( \boldsymbol{A} \cdot \boldsymbol{\nabla'} \right)  \boldsymbol{U} = \frac{\dot R}{R}~\mathscr{P_{\perp}} \boldsymbol{A}~,
\end{equation}
where $\boldsymbol{A}$ is a generic vector field and $\mathscr{P_{\perp}}$ is the projector in the cross-stream direction, i.e. $\mathscr{P_{\perp}} \boldsymbol{A} = \left( A_1, 0, A_3 \right)$. To render more simple our equations we introduce the new velocity field $\boldsymbol{\tilde{u}}(\boldsymbol{x}',t) = \boldsymbol{u'}(\boldsymbol{x}',t) - \boldsymbol{U}(\boldsymbol{x}',t)$
the previous equations can be rewritten as
\begin{equation}
\frac{\partial \rho'}{\partial t} = - \boldsymbol{ \nabla' } \cdot \left( \rho' \boldsymbol{\tilde{u}} \right) -2 \rho' \frac{\dot R}{R}
\end{equation}
\begin{equation} \label{eq:excan}
\frac{\partial \boldsymbol{\tilde{u}}}{\partial t} = \left( \boldsymbol{\mathit{f'}}_{exp} - \frac{\partial \boldsymbol{U}}{\partial t} \right) - \left( \boldsymbol{\tilde{u}} \cdot \boldsymbol{\nabla'} \right) \boldsymbol{\tilde{u}} - \frac{\dot R}{R} \mathscr{P_{\perp}} \boldsymbol{\tilde{u}} - \frac{1}{\rho'} \left[ \boldsymbol{ \nabla' } \left( P' + \frac{\boldsymbol{B'}^2}{2} \right) - \left( \boldsymbol{B'} \cdot \boldsymbol{ \nabla' } \right)   \boldsymbol{B'} \right] 
+ \frac{1}{\rho' \mathcal{R}} \boldsymbol{\nabla'} \cdot \xi'
\end{equation}
\begin{equation}
\frac{\partial \boldsymbol{B'}}{\partial t}  = \boldsymbol{ \nabla' } \times \left( \boldsymbol{\tilde{u}} \times \boldsymbol{B'} \right) - 2\frac{\dot R}{R} \boldsymbol{B'} +\frac{\dot R}{R} \mathscr{P_{\perp}} \boldsymbol{B'}
+ \frac{1}{\mathcal{R}_m} \nabla'^2 \boldsymbol{B'}
\end{equation}
\begin{equation}
\frac{\partial T'}{\partial t} = - \left( \boldsymbol{\tilde{u}} \cdot \boldsymbol{ \nabla' } \right) T'
\end{equation}
In particular the first term between round brackets  in the momentum equation~(\ref{eq:excan}) cancels thanks to the equation~(\ref{eq:exprim}). In this way we have a problem that \emph{numerically} is invariant along the $z$~direction and hence  can be solved with a 2D numerical code.

To simplify further the equations we introduce the average scalings of the physical fields obtained from the matter conservation law and the Alfv\'en theorem. The new fields are the ones with the hat:
\begin{equation}
\rho = \left( \frac{R_0}{R} \right)^2 \hat{\rho}, \quad 
B_x = \left( \frac{R_0}{R} \right)  \hat{B}_x, \quad B_y = \left( \frac{R_0}{R} \right)^2  \hat{B}_y,
\quad B_z = \left( \frac{R_0}{R} \right)  \hat{B}_z,
\end{equation}
and for a matter of convenience we introduce the scalings
\begin{equation}
\tilde{u}_x = \left( \frac{R_0}{R} \right)^{-1} \hat{u}_x, \quad 
\tilde{u}_y = \hat{u}_y, \quad 
\tilde{u}_z = \left( \frac{R_0}{R} \right)^{-1} \hat{u}_z,
\end{equation}
and
\begin{equation}
\boldsymbol{\widehat{ \nabla }} = \left( \frac{\partial}{\partial x'}, \frac{\partial}{\partial y'},
\frac{\partial}{\partial z'} \right)~.
\end{equation}
Notice that for two functions $f$ and $g$ for which holds
\begin{equation}
g = \left( \frac{R_0}{R} \right)^{\alpha} f~,
\end{equation}
for the time derivative it follows
\begin{equation}
\frac{\partial g}{\partial t} = \left( \frac{R_0}{R} \right)^{\alpha} \left[ \frac{\partial f}{\partial t} - \alpha \frac{ \dot{R} }{R} f \right]~.
\end{equation}
Furthermore it can be easily shown that
\begin{equation}
\boldsymbol{ \tilde{u} } \cdot \boldsymbol{\nabla'} = \boldsymbol{\hat{u}} \cdot \boldsymbol{\widehat{\nabla}}~, \qquad
\boldsymbol{\nabla'} \cdot \boldsymbol{\tilde{u}} = \boldsymbol{\widehat{\nabla}} \cdot \boldsymbol{\hat{u}} 
\end{equation}
At last, the equations that we solve with our 2D numerical code are given by (in what follows all the 
fields components should have a hat, that we omit for the sake of simplicity):
\begin{equation} \label{eq:add1}
\frac{\partial \rho}{\partial t}  =  \left( \frac{\partial \rho}{\partial t} \right)_x 
-u_y \frac{\partial \rho}{\partial y} - \rho \frac{\partial u_y}{\partial y},
\end{equation}
\begin{equation} \label{eq:add2}
\frac{\partial u_x}{\partial t} = \left( \frac{\partial u_x}{\partial t} \right)_x -u_y
\frac{\partial u_x}{\partial y} + \left( \frac{R_0}{R} \right)^2 \frac{B_y}{\rho}
\frac{\partial B_x}{\partial y} - \frac{2 \dot{R}}{R} u_x 
+ \frac{1}{\rho \mathcal{R}} \left[ \frac{4}{3}
\frac{\partial^2 u_x}{\partial x^2} +\frac{1}{3} \frac{\partial^2 u_y}{\partial x \partial y}
+ \frac{\partial^2 u_x}{\partial y^2} \right],
\end{equation}
\begin{equation} \label{eq:add3}
\frac{\partial u_y}{\partial t} = \left( \frac{\partial u_y}{\partial t} \right)_x -u_y
\frac{\partial u_y}{\partial y} - 
\left( \frac{\partial T}{\partial y} + \frac{T}{\rho} \frac{\partial \rho}{\partial y} \right)
- \frac{B_x}{\rho} \frac{\partial B_x}{\partial y} - \frac{B_z}{\rho} \frac{\partial B_z}{\partial y} 
+ \frac{1}{\rho \mathcal{R}} \left[ 
\frac{\partial^2 u_y}{\partial x^2} +\frac{1}{3}  \frac{\partial^2 u_x}{\partial x \partial y} + \frac{4}{3} \frac{\partial^2  u_y}{\partial y^2}\right], 
\end{equation}
\begin{equation} \label{eq:add4}
\frac{\partial u_z}{\partial t} = \left( \frac{\partial u_z}{\partial t} \right)_x -u_y
\frac{\partial u_z}{\partial y} + \left( \frac{R_0}{R} \right)^2 \frac{B_y}{\rho}
\frac{\partial B_z}{\partial y} - \frac{2 \dot{R}}{R} u_z +
\frac{1}{\rho \mathcal{R}} \left[ \frac{\partial^2 u_z}{\partial
x^2} +  \frac{\partial^2 u_z}{\partial y^2} \right]  ,
\end{equation}
\begin{equation} \label{eq:add5}
\frac{\partial B_z}{\partial t} = \left( \frac{\partial B_z}{\partial t} \right)_x - B_z
\frac{\partial u_y}{\partial y} + B_y \frac{\partial u_z}{\partial y} - u_y \frac{\partial B_z}{\partial y} + \frac{1}{\mathcal{R}_m} \left[ 
\frac{\partial^2 B_z}{\partial x^2} + \frac{\partial^2 B_z}{\partial y^2} \right], 
\end{equation}
\begin{equation} \label{eq:add6}
\frac{\partial B_y}{\partial t} = \left( \frac{\partial B_y}{\partial t} \right)_x - u_y
\frac{\partial B_y}{\partial y} + \frac{1}{\mathcal{R}_m} \left[ 
\frac{\partial^2 B_y}{\partial x^2} + \frac{\partial^2 B_y}{\partial y^2} \right],
\end{equation}
\begin{equation} \label{eq:add7}
\frac{\partial T}{\partial t} \ \, = \left( \frac{\partial T}{\partial t} \right)_x -u_y \frac{\partial T}{\partial y}
\end{equation}
\begin{equation} \label{eq:add8}
\frac{\partial B_x}{\partial t} = \frac{\partial}{\partial y} \left( u_x B_y - u_y B_x 
\right) + \frac{1}{\mathcal{R}_m} \left[
\frac{\partial^2 B_x}{\partial x^2} + \frac{\partial^2 B_x}{\partial y^2} \right],
\end{equation}
where with the symbol $(\ \ )_x$ we indicate the non-dimensional terms containing only the partial $x$ derivatives:
\begin{equation} \label{eq:ad1}
\left( \frac{\partial \rho}{\partial t} \right)_x \ = -u_x \frac{\partial \rho}{\partial x} - \rho
\frac{\partial u_x}{\partial x}, 
\end{equation}
\begin{equation} \label{eq:ad2}
\left( \frac{\partial u_x}{\partial t} \right)_x = -u_x
\frac{\partial u_x}{\partial x} - \left( \frac{R_0}{R} \right)^{2} \left( \frac{\partial T}{\partial x} +
\frac{T}{\rho} \frac{\partial \rho}{\partial x} \right) 
- \left( \frac{R_0}{R} \right)^2 \frac{1}{\rho} \left( B_z
\frac{\partial B_z}{\partial x} + \left( \frac{R_0}{R} \right)^2 B_y \frac{\partial B_y}{\partial x} \right),
\end{equation}
\begin{equation} \label{eq:ad3} 
\left( \frac{\partial u_y}{\partial t} \right)_x  = - u_x \frac{\partial u_y}{\partial x} +
\left( \frac{R_0}{R} \right)^2  \frac{B_x}{\rho} \frac{\partial B_y}{\partial x}, 
\end{equation}
\begin{equation} \label{eq:ad4}
\left( \frac{\partial u_z}{\partial t} \right)_x = - u_x \frac{\partial u_z}{\partial x} +
\left( \frac{R_0}{R} \right)^2 \frac{B_x}{\rho} \frac{\partial B_z}{\partial x}, 
\end{equation}
\begin{equation} \label{eq:ad5}
\left( \frac{\partial B_z}{\partial t} \right)_x = - B_z \frac{\partial u_x}{\partial x} + B_x \frac{\partial u_z}{\partial x} -u_x \frac{\partial B_z}{\partial x},
\end{equation}
\begin{equation} \label{eq:ad6} 
\left( \frac{\partial B_y}{\partial t} \right)_x = - B_y \frac{\partial u_x}{\partial x} + B_x 
\frac{\partial u_y}{\partial x} - u_x \frac{\partial B_y}{\partial x},
\end{equation}
\begin{equation} \label{eq:ad7}
\left( \frac{\partial T}{\partial t} \right)_x = - u_x \frac{\partial T}{\partial x}
\end{equation}
In equations (\ref{eq:add1})--(\ref{eq:add8}) also the diffusive terms would be affected by expansion factors, but we neglect these factors because in numerical codes diffusive terms play a numerically stabilizing role, which would be affected by such expansion factors. 

From the terms (\ref{eq:ad1})--(\ref{eq:ad7}) one constructs the characteristic polynomial for the $x$ derivative terms, that we use to implement the nonreflecting boundary conditions (see \citet{ein:2001}). This yields the projected entropy, Alfv\'en, fast, slow characteristic speeds for the Expanding Box Model $\lambda$ in non dimensional form as
\begin{equation}
\lambda_0 = u_x, \quad \lambda_{a^{\pm}} = u_x \pm \xi a'_x, \quad
\lambda_{f^{\pm}}= u_x \pm f' , \quad  \lambda_{s^{\pm}} = u_x
\pm s'~, 
\end{equation}
where $\boldsymbol{a'}$ and $c'$ are the modified Alfv\'en and sound speeds
\begin{equation}
c'=  \frac{R_0}{R}~\sqrt{T}~, \qquad 
a'_x = \frac{R_0}{R}~\frac{B_x}{\sqrt{\rho}}~, \qquad 
a'_y = \left( \frac{R_0}{R} \right)^2 \frac{B_y}{\sqrt{\rho}}~, \qquad 
a'_z = \frac{R_0}{R}~\frac{B_z}{\sqrt{\rho}}~,
\end{equation}
and $\xi = a'_x / |a'_x|$.
The fast and slow speeds are defined by:
\begin{equation}
{f'}^2 = \frac{1}{2} \left[ \left( {c'}^2 + {a'}^2 \right) + \sqrt{
\left( {c'}^2 + {a'}^2 \right)^2 - 4 {a'_x}^2 {c'}^2} \ \right],
\end{equation}
\begin{equation}
{s'}^2 = \frac{1}{2} \left[ \left( {c'}^2 + {a'}^2 \right) - \sqrt{
\left( {c'}^2 + {a'}^2 \right)^2 - 4 {a'_x}^2 {c'}^2} \ \right].
\end{equation}
Indicating with $\alpha'_1$, $\alpha'_2$ and $a'_{\perp}$ the quantities
\begin{equation}
\alpha'_1 = \sqrt{\frac{{f'}^2 - {a'_x}^2}{{f'}^2 - {s'}^2}}~, \qquad
\alpha'_2 = \sqrt{\frac{{f'}^2 - {c'}^2}{{f'}^2 - {s'}^2}}~, \qquad
a'_{\perp}= \sqrt{{a'_y}^2+{a'_z}^2}~,
\end{equation}
the projected characteristics are then given by the following expressions: \\ the entropy characteristic,
\begin{equation}
\mathscr{L}_0 = \frac{u_x c'}{T} \frac{\partial T}{\partial x}~,
\end{equation}
the Alfv\'en characteristics,
\begin{equation}
\mathscr{L}_{a^+} = \left(u_x+\xi a'_x\right) \left[
- \frac{R_0}{R} \frac{a'_z}{a'_{\perp}} \frac{\partial u_y}{\partial x} +
 \frac{a'_y}{a'_{\perp}} \frac{\partial u_z}{\partial x} -
 \frac{\xi\frac{R_0}{R}}{\sqrt{\rho}} \left(
 \frac{a'_y}{a'_{\perp}} \frac{\partial B_z}{\partial x} - \frac{a'_z}{a'_{\perp}} \frac{\partial B_y}{\partial x} \right) \right],
\end{equation}
\begin{equation}
\mathscr{L}_{a^-} = \left(u_x-\xi a'_x\right) \left[
- \frac{R_0}{R} \frac{a'_z}{a'_{\perp}} \frac{\partial u_y}{\partial x} +
 \frac{a'_y}{a'_{\perp}} \frac{\partial u_z}{\partial x} +
 \frac{\xi\frac{R_0}{R}}{\sqrt{\rho}} \left(
 \frac{a'_y}{a'_{\perp}} \frac{\partial B_z}{\partial x} - \frac{a'_z}{a'_{\perp}} \frac{\partial B_y}{\partial x} \right) \right],
\end{equation}
the slow mode characteristics
\begin{multline}
\mathscr{L}_{s^+} = \left( u_x+s' \right)
\Bigg\{-\alpha'_2 \left[ \left(\frac{R_0}{R}\right)^{2}
\frac{1}{c'} \left(
\frac{T}{\rho} \frac{\partial \rho}{\partial x} + \frac{\partial T}{\partial x} \right) +\xi \frac{a'_x}{f'} \frac{\partial u_x}{\partial x}
\right] + \alpha'_1 \Bigg[ \frac{c'}{f'\sqrt{\rho}}~\cdot \\ 
\cdot \left( \frac{R_0}{R}
\frac{a'_z}{a'_{\perp}} \frac{\partial B_z}{\partial x} + 
\left(\frac{R_0}{R}\right)^2 \frac{a'_y}{a'_{\perp}} \frac{\partial B_y}{\partial x} \right)
- \xi \left(\frac{a'_z}{a'_{\perp}} \frac{\partial u_z}{\partial x} + \frac{R_0}{R}
  \frac{a'_y}{a'_{\perp}} \frac{\partial u_y}{\partial x} \right) \Bigg] \Bigg\},
\end{multline}
\begin{multline}
\mathscr{L}_{s^-} = \left(u_x-s' \right)
\Bigg\{-\alpha'_2 \left[ \left(\frac{R_0}{R}\right)^{2}
  \frac{1}{c'} \left( 
\frac{T}{\rho} \frac{\partial \rho}{\partial x} + \frac{\partial T}{\partial x} \right) -\xi \frac{a'_x}{f'} \frac{\partial u_x}{\partial x}
\right] + \alpha'_1 \Bigg[ \frac{c'}{f'\sqrt{\rho}}~\cdot \\ 
\cdot \left( \frac{R_0}{R}
\frac{a'_z}{a'_{\perp}} \frac{\partial B_z}{\partial x} + \left(\frac{R_0}{R}\right)^2
\frac{a'_y}{a'_{\perp}} \frac{\partial B_y}{\partial x} \right) + \xi
\left(\frac{a'_z}{a'_{\perp}} \frac{\partial u_z}{\partial x} + \frac{R_0}{R}
 \frac{a'_y}{a'_{\perp}} \frac{\partial u_y}{\partial x} \right) \Bigg] \Bigg\},
\end{multline}
and the fast mode characteristics
\begin{multline}
\mathscr{L}_{f^+} = \left(u_x+f'\right)
\Bigg\{ \alpha'_1 \left[ \left(\frac{R_0}{R}\right)^{2}
\frac{1}{f'} \left( 
\frac{T}{\rho} \frac{\partial \rho}{\partial x} + \frac{\partial T}{\partial x} \right) + \frac{\partial u_x}{\partial x}
\right] + \alpha'_2 \Bigg[ \frac{1}{\sqrt{\rho}}~\cdot  \\
\cdot \left( \frac{R_0}{R} \frac{a'_z}{a'_{\perp}}
\frac{\partial B_z}{\partial x} + \left(\frac{R_0}{R}\right)^2
\frac{a'_y}{a'_{\perp}} \frac{\partial B_y}{\partial x} \right) - \frac{a'_x}{f'} 
\left(\frac{a'_z}{a'_{\perp}} \frac{\partial u_z}{\partial x} + \frac{R_0}{R}
\frac{a'_y}{a'_{\perp}} \frac{\partial u_y}{\partial x} \right) \Bigg] \Bigg\},
\end{multline}
\begin{multline}
\mathscr{L}_{f^-} = \left(u_x-f'\right)
\Bigg\{ \alpha'_1 \left[ \left(\frac{R_0}{R}\right)^{2}
\frac{1}{f'} \left( 
\frac{T}{\rho} \frac{\partial \rho}{\partial x} + \frac{\partial T}{\partial x} \right) - \frac{\partial u_x}{\partial x}
\right] + \alpha'_2 \Bigg[ \frac{1}{\sqrt{\rho}}~\cdot   \\
\cdot \left( \frac{R_0}{R} \frac{a'_z}{a'_{\perp}}
\frac{\partial B_z}{\partial x} + \left(\frac{R_0}{R}\right)^2
\frac{a'_y}{a'_{\perp}} \frac{\partial B_y}{\partial x} \right) + \frac{a'_x}{f'} 
\left(\frac{a'_z}{a'_{\perp}} \frac{\partial u_z}{\partial x} + \frac{R_0}{R}
\frac{a'_y}{a'_{\perp}} \frac{\partial u_y}{\partial x} \right) \Bigg] \Bigg\}.
\end{multline}
In terms of characteristics, the time derivatives (\ref{eq:ad1})--(\ref{eq:ad7}) are then given explicitly by:
\begin{equation} \label{eq:nr1}
\left( \frac{\partial \rho}{\partial t} \right)_x = \frac{\rho}{c'} \mathscr{L}_0 + \rho
\Bigg[ \frac{ \alpha'_2 }{c'} \left(
\frac{\mathscr{L}_{s^+} +\mathscr{L}_{s^-}}{2} \right) -\frac{\alpha'_1}{f'}
\left( \frac{\mathscr{L}_{f^+} + \mathscr{L}_{f^-}}{2} \right)
\Bigg],
\end{equation}
\begin{gather}
\left( \frac{\partial u_x}{\partial t} \right)_x = 
\frac{s'}{c'} \alpha'_2 \left(
\frac{\mathscr{L}_{s^+} - \mathscr{L}_{s^-}}{2} \right) -
\alpha'_1 
 \left( \frac{\mathscr{L}_{f^+} - \mathscr{L}_{f^-}}{2} \right),
\end{gather}
\begin{equation}
\left( \frac{\partial u_y}{\partial t} \right)_x = \frac{R}{R_0} \frac{a'_z}{a'_{\perp}}
\left( \frac{\mathscr{L}_{a^+} +\mathscr{L}_{a^-}}{2} \right) + 
\frac{R}{R_0} \frac{a'_y}{a'_{\perp}} \Bigg[ \xi
\alpha'_1 \left (
\frac{\mathscr{L}_{s^+} - \mathscr{L}_{s^-}}{2} \right) 
+ \frac{a'_x}{f'} \alpha'_2
\left( \frac{\mathscr{L}_{f^+} - \mathscr{L}_{f^-}}{2} \right)
\Bigg], 
\end{equation}
\begin{equation}
\left( \frac{\partial u_z}{\partial t} \right)_x = - \frac{a'_y}{a'_{\perp}} \left(
\frac{\mathscr{L}_{a^+} +\mathscr{L}_{a^-}}{2} \right) + 
\frac{a'_z}{a'_{\perp}} \Bigg[ \xi
\alpha'_1 \left (
\frac{\mathscr{L}_{s^+} - \mathscr{L}_{s^-}}{2} \right) 
+ \frac{a'_x}{f'} \alpha'_2 
\left( \frac{\mathscr{L}_{f^+} - \mathscr{L}_{f^-}}{2} \right)
\Bigg], 
\end{equation}
\begin{equation}
\left( \frac{\partial B_z}{\partial t} \right)_x = \xi  \frac{R}{R_0} \sqrt{\rho}
\frac{a'_y}{a'_{\perp}} 
\left( \frac{\mathscr{L}_{a^+} - \mathscr{L}_{a^-}}{2} \right) -
\frac{R}{R_0} \sqrt{\rho} \frac{a'_z}{a'_{\perp}}
 \cdot \Bigg[ \frac{c'}{f'} \alpha'_1 
\left ( \frac{\mathscr{L}_{s^+} +\mathscr{L}_{s^-}}{2} \right)
+ \alpha'_2  \left( \frac{\mathscr{L}_{f^+} + \mathscr{L}_{f^-}}{2}
\right) \Bigg], 
\end{equation}
\begin{equation}
\left( \frac{\partial B_y}{\partial t} \right)_x = - \xi \left( \frac{R}{R_0} \right)^2
\sqrt{\rho} \frac{a'_z}{a'_{\perp}}
\left( \frac{\mathscr{L}_{a^+} - \mathscr{L}_{a^-}}{2} \right) -
\left( \frac{R}{R_0} \right)^2 \sqrt{\rho} \frac{a'_y}{a'_{\perp}}
\cdot \Bigg[ \frac{c'}{f'} \alpha'_1 
\left ( \frac{\mathscr{L}_{s^+} +\mathscr{L}_{s^-}}{2} \right) 
+ \alpha'_2
\left( \frac{\mathscr{L}_{f^+} + \mathscr{L}_{f^-}}{2} \right) \Bigg],
\end{equation}
\begin{equation} \label{eq:nr7}
\left( \frac{\partial T}{\partial t} \right)_x = - \frac{T}{c'} \mathscr{L}_0
\end{equation}
Nonreflecting boundary conditions are obtained by setting to zero the values of the characteristics for the inward propagating waves in equations~(\ref{eq:nr1})--(\ref{eq:nr7}) (see \citet{thom:1987}, \citet{thom:1990}, \citet{vana:1989} and \citet{roe:1996}).




\clearpage



\clearpage

\clearpage



\begin{thebibliography}{}
\bibitem[Chandrasekhar~(1961)]{chandra} Chandrasekhar, S., Hydrodynamic and Hydromagnetic Stability, Cambridge University Press, Cambridge, 1961
\bibitem[Dahlburg et~al.~(1998)]{dbe:1998} Dahlburg, R.~B.,
Boncinelli, P., \& Einaudi, G. 1998 Phys. Plasmas, 5, 79
\bibitem[Dahlburg~(1998)]{dahl:1998} Dahlburg, R.~B., 1998 Phys. Plasmas, 5, 133
\bibitem[Einaudi et~al.~(1999)]{einaudi:1999} Einaudi, G., Boncinelli,
P., Dahlburg, R.~B., \& Karpen, J.~T., 1999 J.  Geophys. Res., 104, 521
\bibitem[Einaudi et~al.~(2001)]{ein:2001} Einaudi, G., Chibbaro, S.,
Dahlburg, R.~B., \& Velli, M. 2001 Astrophys. J., 547, 1167
\bibitem[Ferrari and Trussoni~(1983)]{ferr:1983} Ferrari, A., \& Trussoni, E., 1983 MNRAS, 205, 515
\bibitem[Guhathakurta et~al.~(1996)]{guha:1996} Guhathakurta, M., Holzer, T.E., 
\& MacQueen, R.M. 1996 \apj, 458, 817
\bibitem[Grappin and Velli~(1996)]{grap:1996} Grappin, R., \& Velli, M., 1996 J. Geophys. 
Res., 101, 425
\bibitem[Grappin et~al.~(1993)]{grap:1993} Grappin, R.,  Velli, M., \& Mangeney A. 1993 Phys. Rev. Lett., 70, 2190
\bibitem[Gosling et~al.~(1981)]{gosling} Gosling, J.T., Borrini, G., Asbridge, J.R., Bame, S.J., Feldman, W. C., \& Hansen, R.T., 1981 J. Geophys. Res., 86, 5438
\bibitem[Lele~(1992)]{lele:1992} Lele, S.~K., 1992 J. Comput. Phys., 103, 16
\bibitem[Parker~(1954)]{park:1954} Parker, E.~N., 1954 Phys. Rev., 96, 1686
\bibitem[Parker~(1957)]{park:1957} Parker, E.~N., 1957 \apjs, 3, 51
\bibitem[Pneuman and Cargill~(1985)]{pc:1985} Pneuman, G.~W., \&
Cargill, P.~J., 1985 \apj, 288, 653
\bibitem[Ray~(1982)]{ray:1982} Ray, T.P., 1982 MNRAS, 198, 617
\bibitem[Roe and Balsara~(1996)]{roe:1996} Roe, P.L., \& Balsara, D.S., 1996 SIAM J. Appl. Math., 56, 57
\bibitem[Schmidt and Cargill~(2000)]{sc:2000} Schmidt, J.~M., \&
Cargill, P.~J., 2000 J. Geophys. Res., 105, 10455
\bibitem[Sheeley et~al.~(1997)]{sheeley:1997} Sheeley, N.~R. et al., 1997 \apj, 484, 472
\bibitem[Stribling et~al.~(1996)]{stribling:1996} Stribling, T., Roberts, D.~A., \& Goldstein, M.~L.,
1996 J.  Geophys. Res., 101, 27603
\bibitem[Thompson~(1987)]{thom:1987} Thompson, K.~W., 1987 J. Comput. Phys., 68, 1
\bibitem[Thompson~(1990)]{thom:1990} Thompson, K.~W., 1990 J. Comput. Phys., 89, 439
\bibitem[Vanajakshi et~al.~(1989)]{vana:1989} Vanajakshi, T.~X.,
Thompson, K.~W., \& Black, ~D.C., 1989 J. Comput. Phys., 84, 343
\bibitem[Wang et~al.~(1998)]{wang:1998} Wang, Y.-M. et al., 1998 \apjl, 498, L165
\end{thebibliography}
\end{document}